\documentclass[runningheads,orivec,a4paper]{llncs}
\usepackage[utf8]{inputenc}
\usepackage{amsmath,amssymb,amsfonts}
\usepackage{dsfont}
\usepackage{listings}
\usepackage{graphicx}
\usepackage[dvipsnames]{xcolor}
\usepackage{amsopn}
\usepackage{tabularx,booktabs,subcaption}
\usepackage{hyperref}
\usepackage{array}
\usepackage{upgreek}

\newcolumntype{Y}{>{\centering\arraybackslash}X}
\newcolumntype{C}[1]{>{\centering\let\newline\\\arraybackslash\hspace{0pt}}m{#1}}

\usepackage{accents}
\newcommand{\vardbtilde}[1]{\tilde{\raisebox{0pt}[0.85\height]{$\tilde{#1}$}}}

\def\pset{{P\left(\mathbf{L}\right)^N}}
\def\gset{{\left(2^{\mathbf{L}}\right)^N}}

\newcommand{\uzl}{University Hospital Leuven}

\newcommand{\subheading}[1]{\textbf{#1}}

\title{Label-set Loss Functions for Partial Supervision: Application to Fetal Brain 3D MRI Parcellation}

\titlerunning{Label-set Loss Functions for Partial Supervision}

\author{Lucas Fidon\inst{1}
    \and Michael Aertsen\inst{2}
    \and Doaa Emam\inst{4,5}
    \and Nada Mufti\inst{1,3,4}
    \and \\Fr\'ed\'eric Guffens\inst{2}
    \and Thomas Deprest\inst{2}
    \and Philippe Demaerel\inst{2}
    \and \\Anna L. David\inst{3,4}
    \and Andrew Melbourne\inst{1}
	\and S\'ebastien Ourselin\inst{1}
	\and \\Jan Deprest\inst{2,3,4}
	\and Tom Vercauteren\inst{1}}

\authorrunning{Lucas Fidon et al.}

\institute{
School of Biomedical Engineering \& Imaging Sciences, King's College London, UK 
\and
Department of Radiology, University Hospitals Leuven, Belgium
\and
Institute for Women's Health, University College London, UK
\and
Department of Obstetrics and Gynaecology, University Hospitals Leuven, Belgium
\and
Department of Gynecology and Obstetrics, University Hospitals Tanta, Egypt
}

\begin{document}

\maketitle

\begin{abstract}
    Deep neural networks have increased the accuracy of automatic segmentation,
    however their accuracy depends on the availability of a large number of fully segmented images.
    %
    Methods to train deep neural networks using images for which some, but not all, regions of interest are segmented are necessary to make better use of partially annotated datasets.
    In this paper, we propose the first axiomatic definition of label-set loss functions that are the loss functions that can handle partially segmented images.
    We prove that there is one and only one method to convert a classical loss function for fully segmented images into a proper label-set loss function.
    %
    Our theory also allows us to define the leaf-Dice loss, a label-set generalisation of the Dice loss particularly suited for partial supervision with only \emph{missing} labels.
    Using the 
    leaf-Dice loss, we set a new state of the art in partially supervised learning for fetal brain 3D 
    MRI 
    segmentation.
    %
    We achieve 
    a
    deep neural network able to segment white matter, ventricles, cerebellum, extra-ventricular CSF, cortical gray matter, deep gray matter, brainstem, and corpus callosum based 
    on fetal brain 3D MRI 
    of anatomically normal fetuses or with open spina bifida.
    Our implementation of the proposed label-set loss functions is available at \url{https://github.com/LucasFidon/label-set-loss-functions}
\end{abstract}

\section{Introduction}
The parcellation of fetal brain MRI is essential for 
the study of fetal brain development~\cite{benkarim2017toward}.
Reliable analysis and evaluation of fetal brain structures
could also support diagnosis of central nervous system pathology, patient selection for fetal surgery, evaluation and prediction of outcome, hence also parental counselling~\cite{aertsen2019reliability,danzer2020fetal,moise2016current,sacco2019fetal,zarutskie2019prenatal}.
Deep learning sets the state of the art for the automatic parcellation of fetal brain MRI~\cite{fetit2020deep,khalili2019automatic,payette2020efficient,payette2019longitudinal}.
Training a deep learning model requires a large amount of accurately annotated data.
However, manual parcellation of fetal brain 3D MRI requires highly skilled raters and is time-consuming.


Training a deep neural network for segmentation with partially segmented images is known as partially supervised learning~\cite{zhou2019prior}.
Recent studies have proposed to use partially supervised learning for body segmentation in CT~\cite{dmitriev2019learning,fang2020multi,shi2021marginal,zhou2019prior} and for the joined segmentation of brain tissues and lesions in MRI~\cite{dorent2021learning,roulet2019joint}.
One of the main challenges in partially supervised learning is to define loss functions that can handle partially segmented images.
Several previous studies have proposed to adapt existing loss functions for fully supervised learning using a somewhat ad hoc marginalization method~\cite{fang2020multi,roulet2019joint,shi2021marginal}. Theoretical motivations for 
such marginalisation were missing.
It also
remains unclear whether it
is the only way to build loss functions for partially supervised learning.

\begin{figure}[bt]
    \centering
    \includegraphics[width=\textwidth, trim={0.1cm 5.12cm 0.1cm 5.8cm},clip]{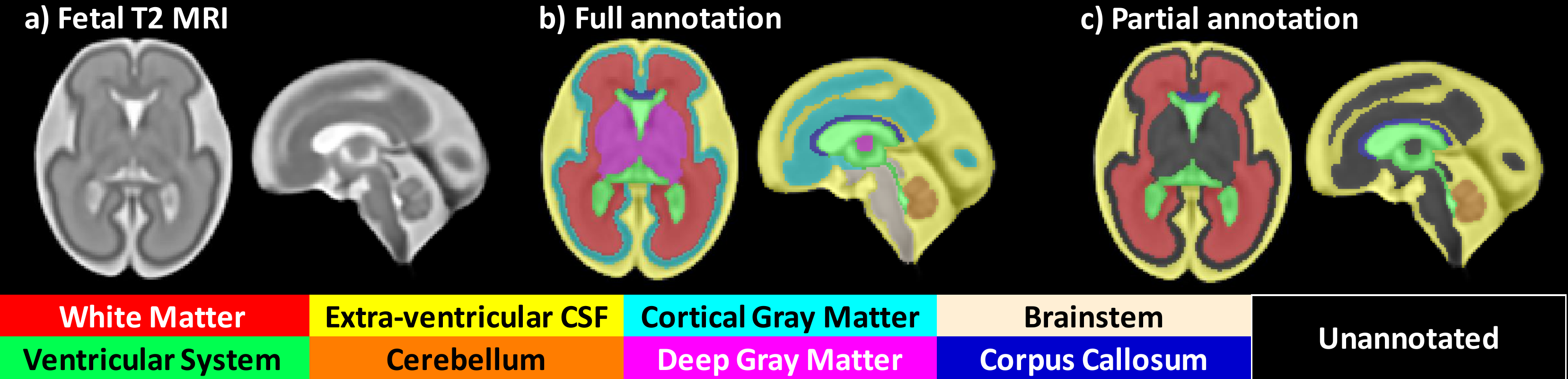}
    \caption{
    Illustration of partial annotations on a control fetal brain MRI~\cite{gholipour2017normative}.
    b) all the leaf-labels are annotated.
    c) partial annotations where cortical gray matter (CGM), deep gray matter (DGM), and brainstem (B) are not annotated.
    In this cases, unannotated voxels have the label-set annotation \{CGM, DGM, B\}.
    }
    \label{fig:partial_annotations}
\end{figure}

In this paper, we give the first theoretical framework for loss functions in partially supervised learning.
We call those losses \emph{label-set loss functions}.
While in a fully supervised scenario, each voxel is assigned a single label, which we refer to as a \emph{leaf-label} hereafter to avoid ambiguity;
with partial supervision, each voxel is assigned a combined label, which we refer to as a \emph{label-set}.
%
As illustrated in Fig.~\ref{fig:partial_annotations}, a typical example of partial supervision arises when there are missing leaf-label annotations.
In which case the regions
that were not segmented manually are grouped under one label-set (\emph{unannotated}).
%
Our theoretical contributions are threefold:
1) We introduce an axiom that label-set loss functions must satisfy to guarantee 
compatibility across label-sets and leaf-labels;
2) we propose a generalization of the Dice loss, leaf-Dice, that satisfies our axiom for the common case of missing leaf-labels; and
3) we demonstrate that there is one and only one way to convert a classical segmentation loss for fully supervised learning into a loss function for partially supervised learning that complies with our axiom.
This theoretically justifies the marginalization method used in previous work~\cite{fang2020multi,roulet2019joint,shi2021marginal}.

In our experiments, we propose the first application of
partial supervision
to fetal brain 3D MRI segmentation.
We use
$244$ fetal brain volumes from 3 clinical centers, to evaluate the automatic segmentation of 8 tissue types for both normal fetuses and fetuses with open spina bifida.
We compare the proposed leaf-Dice to another labels-set loss~\cite{shi2021marginal} and to two other baselines.
Our results support the superiority of labels-set losses that comply with the proposed axiom and show that the leaf-Dice loss significantly outperforms the three other methods.

\section{Theory of Label-set Loss Functions}

In fully-supervised learning, we learn from
ground-truth
segmentations $\textbf{g} \in \mathbf{L}^N$
where $N$ is the number of voxels and $\mathbf{L}$ is the set of 
final labels
(e.g. white matter, 
ventricular system).
We denote elements of $\mathbf{L}$ as \emph{leaf-labels}.
In contrast,
with partial supervision,
$\textbf{g} \in \left(2^{\mathbf{L}}\right)^N$, where $2^{\mathbf{L}}$ is the set of subsets of $\mathbf{L}$.
In other words, 
each voxel 
annotation is
a \emph{set} of leaf-labels 
(a \emph{label-set}).
The label-set of an annotated white matter voxel would be simply \{white matter\}, and the label-set of 
a
voxel that could be either white matter or deep gray matter would be 
\{white matter, deep gray matter\}.
%
In both fully-supervised and partially-supervised learning,
the network is trained to
perform full segmentation predictions: 
$\textbf{p}\in\pset$, where $P(\mathbf{L})$ is the space of probability vectors for leaf-labels.

\subsection{Label-set Loss Functions}\label{sec:axiom}

A loss function 
$\mathcal{L}_{partial}(\cdot,\cdot)$
for partially supervised learning can be any differentiable function
that compares a proposed probabilistic network output for leaf-labels, $\textbf{p} \in \pset$, and a partial label-set ground-truth annotation, $\textbf{g} \in \gset$,
\begin{equation}
\begin{split}
    \mathcal{L}_{partial}:\,\,& \pset \times \gset \xrightarrow{} \mathds{R}\\
\end{split}
\end{equation}
However, such $\mathcal{L}_{partial}$, in general, may consider all possible label-sets as independent and ignore the relationship between a label-set and its constituent leaf-labels.
We claim that segmentation loss functions for partially supervised learning must be compatible with the semantic of label-set inclusion.
For example, in the case of three leaf-labels $\textbf{L}=\{l_1, l_2, l_3\}$, let voxel $i$ be labeled with the label-set $g_i = \{l_1, l_2\}$. 
%
We know that the true leaf-label is either $l_1$ or $l_2$.
Therefore the exemplar leaf-label probability vectors $\textbf{p}_i = (0.4, 0.4, 0.2)$, $\textbf{q}_i = (0.8, 0, 0.2)$, and $\textbf{h}_i = (0, 0.8, 0.2)$ need to be equivalent conditionally to the ground-truth partial annotation $g_i$.
That is, the value of the loss function should be the same whether the predicted leaf-label probability vector for voxel $i$ is $\textbf{p}_i$, $\textbf{q}_i$, or $\textbf{h}_i$.

Formally, let us define the marginalization function $\Phi$
as 
\begin{equation*}
    \begin{split}
        \Phi: \,\,& P\left(\mathbf{L}\right)^N \times \left(2^{\mathbf{L}}\right)^N 
        \xrightarrow{} P\left(\mathbf{L}\right)^N\\
         & (\textbf{p}, \textbf{g}) \mapsto
         (\tilde{p}_{i,c})
    \end{split}
     \,\,\textup{s.t.}\quad \forall i,c,\,
         \left\{
            \begin{array}{cc}
              \tilde{p}_{i,c} = \frac{1}{|g_i|} \sum_{c' \in g_i} p_{i,c'}& \textup{if}\,\, c \in g_i\\
              \tilde{p}_{i,c} = p_{i,c} &  \textup{if}\,\, c \not \in g_i
            \end{array}
         \right.
\end{equation*}
In the previous example, 
$
\Phi(\textbf{p};\textbf{g})_i
= \Phi(\textbf{q};\textbf{g})_i
= \Phi(\textbf{h};\textbf{g})_i
= (0.4, 0.4, 0.2).
$
We define \textbf{label-set loss functions} as the functions $\mathcal{L}_{partial}$ that satisfy the axiom
\begin{equation}
    \forall \textbf{g},
    \forall \textbf{p},\textbf{q},\quad
    \Phi(\textbf{p}; \textbf{g}) = \Phi(\textbf{q}; \textbf{g})
    \implies 
    \mathcal{L}_{partial}(\textbf{p}, \textbf{g})
    = \mathcal{L}_{partial}(\textbf{q}, \textbf{g})
    \label{eq:axiom}
\end{equation}
We demonstrate that a loss $\mathcal{L}$ is a label-set loss function if and only if
\begin{equation}
    \forall (\textbf{p},\textbf{g}), \quad \mathcal{L}(\textbf{p}, \textbf{g}) = \mathcal{L}\left(\Phi(\textbf{p}; \textbf{g}), \textbf{g}\right)
    \label{eq:lemma_general}
\end{equation}
See the supplementary material for a proof of this equivalence.

\subsection{Leaf-Dice: A Label-set Generalization of the Dice Loss}\label{sec:leafdice}

In this section, as per previous 
work~\cite{dmitriev2019learning,dorent2021learning,fang2020multi,roulet2019joint,shi2021marginal,zhou2019prior}, we consider the particular case in which, per training example,
there is only one label-set 
that is not a singleton and contains all the 
leaf-labels that were not manually segmented in this example.
An illustration for fetal brain segmentation can be found in Fig.\ref{fig:partial_annotations}.

We propose a generalization of the mean class Dice Loss~\cite{fidon2017generalised,milletari2016v} for this particular case
and prove that it satisfies our axiom~\eqref{eq:axiom}.

Let $\textbf{g} \in \gset$ be a partial label-set segmentation such that there exists a label-set $\mathbf{L}_{\textbf{g}}' \subsetneq \mathbf{L}$ that contains all the leaf-labels that were not manually segmented for 
this subject.
%
Therefore, $\textbf{g}$ takes its values in 
$\{\mathbf{L}_{\textbf{g}}'\} \cup \left\{\{c\}\,|\,c \in \mathbf{L}\setminus\mathbf{L}_{\textbf{g}}'\right\}$.
We demonstrate that the leaf-Dice loss defined below is a label-set loss function 
%
\begin{equation}
    \forall \textbf{p},\quad
    \mathcal{L}_{Leaf-Dice}(\textbf{p}, \textbf{g}) =
    1 - 
    \frac{1}{|\mathbf{L}|} 
    \sum_{c \in \mathbf{L}} 
    \frac{2 \sum_i \mathds{1}(g_i =\{c\})\,p_{i,c}}{
    \sum_i \mathds{1}(g_i =\{c\})^{\alpha}
    + \sum_{i} p_{i,c}^{\alpha}
    +\epsilon
    }
    \label{eq:ls_dice}
\end{equation}
where $\alpha \in \{1,2\}$
(in line with the variants of soft Dice encountered in practice), 
and
$\epsilon > 0$ is a small constant.
%
A proof that $\mathcal{L}_{Leaf-Dice}$ satisfies~\eqref{eq:lemma_general} can be found in the supplementary material.
It is worth noting that using
$\mathcal{L}_{Leaf-Dice}$ is not equivalent to just masking out the unannotated voxels,
i.e. the voxels
$i$ such that $g_i=\mathbf{L'}_{\textbf{g}}$.
Indeed,
for all the $c \in \mathbf{L}\setminus \mathbf{L}_{\textbf{g}}'$,
the term $\sum_i p_{i,c}^{\alpha}$ in the denominator pushes $p_{i,c}$ towards $0$ for all the voxels indices $i$ including the indices $i$ for which $g_i=\mathbf{L}_{\textbf{g}}'$.
As a result, when $g_i=\mathbf{L}_{\textbf{g}}'$, only the $p_{i,c}^{\alpha}$ for $c \not \in \mathbf{L}_{\textbf{g}}'$ are pushed toward $0$, which in return pushes the $p_{i,c}^{\alpha}$ for $c \in \mathbf{L}_{\textbf{g}}'$ towards $1$ since $\sum_{c \in \mathbf{L}} p_{i,c}=1$.

\subsection{Converting Classical Loss Functions to Label-set Loss Functions}\label{sec:conversion}
In this section, we demonstrate that there is one and only one canonical method to convert a segmentation loss function for fully supervised learning into a label-set loss function for partially supervised learning satisfying~\eqref{eq:lemma_general}.

Formally, a label-set segmentation $\textbf{g}$ can be seen as a soft segmentation using an injective function
$
\Psi: \left(2^{\mathbf{L}}\right)^N \hookrightarrow{} P\left(\mathbf{L}\right)^N
$
%
that satisfies the relation
\begin{equation}
    \forall \textbf{g} \in \gset,\,\forall i,c,\quad 
        [\Psi(\textbf{g})]_{i,c} > 0 \implies c \in g_i
\end{equation}
Note however, that the function $\Psi$ is not unique. 
Following the maximum entropy principle leads to choose
\begin{equation}
    \Psi_0: (g_i) \mapsto (\tilde{p}_{i,c})
    \quad \textup{s.t.}\quad \forall i,c\,
         \left\{
            \begin{array}{cc}
              \tilde{p}_{i,c} = \frac{1}{|g_i|} & \textup{if}\,\, c \in g_i\\
              \tilde{p}_{i,c} = 0 &  \textup{otherwise}
            \end{array}
         \right.
    \label{eq:psi0}
\end{equation}
We are interested in converting a loss function for fully-supervised learning
$
\mathcal{L}_{fully}:\,\, P\left(\mathbf{L}\right)^N \times P\left(\mathbf{L}\right)^N \xrightarrow{} \mathds{R}
$
into a label-set loss function for partial supervision defined as
$\mathcal{L}_{partial}(\textbf{p}, \textbf{g}) = \mathcal{L}_{fully}(\textbf{p}, \Psi\left(\textbf{g}\right))$.

%
Assuming that $\mathcal{L}_{fully}$ is minimal if and only if the predicted segmentation and the soft ground-truth segmentation are equal,
%
%
we demonstrate in the supplementary material that $ \mathcal{L}_{partial}$ is a label-set loss function if and only if
\begin{equation}
\forall (\textbf{p}, \textbf{g})\in \pset \times \gset,\quad
\left\{
\begin{aligned}
    \mathcal{L}_{partial}(\textbf{p}, \textbf{g})
    &= \mathcal{L}_{fully}\left(\Phi(\textbf{p}; \textbf{g}), \Psi(\textbf{g})\right)\\
    %
    \Psi(\textbf{g}) &= \Psi_0(\textbf{g})
\end{aligned}
\right.
\end{equation}
%
%
Therefore, 
the only way to convert a fully-supervised loss to a loss for partially-supervised learning that complies with our axiom \eqref{eq:axiom} 
is to use the marginalization function $\Phi$ on the predicted segmentation and $\Psi_0$ on the ground-truth partial segmentation.
For clarity, we emphasise that the Leaf-Dice is a generalisation rather than a conversion of the Dice loss.

\subheading{Related Work.}
When $\mathcal{L}_{fully}$ is the mean class Dice loss~\cite{fidon2017generalised,milletari2016v} and the values of $\textbf{g}$ are a partition of $\mathbf{L}$,
like in section~\ref{sec:leafdice},
one can prove that $\mathcal{L}_{partial}$ is the marginal Dice loss~\cite{shi2021marginal} (proof in supplementary material).
Similarly, when $\mathcal{L}_{fully}$ is the cross entropy loss, $\mathcal{L}_{partial}$ is the marginal cross entropy loss~\cite{fang2020multi,roulet2019joint,shi2021marginal}.
Note that in~\cite{fang2020multi,roulet2019joint,shi2021marginal}, the marginalization approach is proposed as a possible method to convert the loss function.
We prove that this is the only method.

\section{Fetal Brain 3D MRI Data with Partial Segmentations}\label{sec:dataset}

In this section, we describe the fetal brain 3D MRI datasets that were used.
%
%

\begin{table}[bt]
    \caption{
	\textbf{
	Number of 3D MRI and number of manual segmentations available per tissue types.}
	\textcolor{red}{\bf WM}: white matter,
	\textcolor{ForestGreen}{\bf Vent}: ventricular system,
	\textcolor{orange}{\bf Cer}: cerebellum,
	\textcolor{Goldenrod}{\bf ECSF}: extra-ventricular CSF,
	\textcolor{Cyan}{\bf CGM}: cortical gray matter,
	\textcolor{Purple}{\bf DGM}: deep gray matter,
	\textcolor{Gray}{\bf BS}: brainstem,
	\textcolor{blue}{\bf CC}: corpus callosum.}
    \begin{tabularx}{\textwidth}{ c c *{9}{Y}}
		\toprule
		\textbf{Train/Test} &
		\textbf{Condition} &
		\textbf{MRI} & 
		\textcolor{red}{\bf WM} & \textcolor{ForestGreen}{\bf Vent} &
		\textcolor{orange}{\bf Cer} & \textcolor{Goldenrod}{\bf ECSF} &
		\textcolor{Cyan}{\bf CGM} & \textcolor{Purple}{\bf DGM} &
		\textcolor{Gray}{\bf BS} & \textcolor{blue}{\bf CC}\\
		\midrule
		Training & Atlas~\cite{gholipour2017normative} &
		18 & 18 & 18 & 18 & 18 & 18 & 18 & 18 & 18\\
		Training & Controls &
		116 & 116 & 116 & 116 & 54 & 0 & 0 & 0 & 18\\
		Training & Spina Bifida & 
		30 &  30 &  30 &  30 & 0  & 0  & 0  & 0  & 0 \\
	    \midrule
	    Testing & Controls & 34 & 34 & 34 & 34 & 34 & 15 & 15 & 15 & 0\\
		Testing & Spina Bifida & 66 & 66 & 66 & 66 & 66 & 25 & 25 & 25 & 41\\
	    \bottomrule
	\end{tabularx}
    \label{tab:data}
\end{table}

\subheading{Training Data for Fully Supervised Learning:}
$18$ fully-annotated control fetal brain MRI from a spatio-temporal~\cite{gholipour2017normative}.

\subheading{Training Data for Partially Supervised Learning:}
$18$ fully-annotated volumes from the fully-supervised dataset, combined with
$146$ partially annotated fetal brain MRI from a private dataset.
The 
segmentations available for those 3D MRI are detailed in Table~\ref{tab:data}.

\subheading{Multi-centric Testing Data:}
$100$ fetal brain 3D MRI.
%
This includes $60$ volumes from \uzl{} and $40$ volumes from the publicly available FeTA dataset~\cite{payette2020comparison}.
The segmentations available for those 3D MRI are detailed in Table~\ref{tab:data}.
%
%
The 3D MRI of the FeTA dataset come from a different center than the training data.
%
%
Automatic brain masks for the FeTA data were obtained using atlas affine registration with a normal fetal brain and a spina bifida spatio-temporal atlas~\cite{fidon2021atlas,gholipour2017normative}.

\subheading{
Image Acquisition and Preprocessing for the Private Dataset.
}
All images were part of routine clinical care and were acquired at \uzl{}.
%
In total, 74 cases with open spina bifida and 135 cases with normal brains, referred as controls, were included.
Three spina bifida cases have been excluded by a pediatric radiologist because 
the quality of the 2D MRI resulted in corrupted 3D MRI which did not allow accurate segmentation.
%
The gestational age at MRI ranged from $20$ weeks to $35$ weeks (median=$26.9$ weeks, IQR=$3.4$ weeks).
For each study, at least three orthogonal T2-weighted HASTE series of the fetal brain were collected on a $1.5$T scanner using an echo time of $133$ms, a repetition time of $1000$ms, with no slice overlap nor gap, pixel size $0.39$mm to $1.48$mm, and slice thickness $2.50$mm to $4.40$mm.
A radiologist attended all the acquisitions for quality control.
The fetal brain 3D MRI were obtained using \texttt{NiftyMIC}~\cite{ebner2020automated} 
a state-of-the-art super resolution and reconstruction algorithm. The volumes were all reconstructed to a resolution of $0.8$ mm isotropic and registered to a standard clinical view.
\texttt{NiftyMIC} also outputs brain masks that were used to define the label-sets and to mask the background~\cite{ranzini2021monaifbs}.

\subheading{Labelling Protocol.}
The labelling protocol is the same as in~\cite{payette2020comparison}.
%
%
The different tissue types were segmented by a trained obstetrician and medical students 
under the supervision of
a paediatric radiologist specialized in fetal brain anatomy, who quality controlled and corrected all manual segmentations.
%
The voxels inside the brain mask that were not annotated by experts were assigned to the label-set containing all the tissue types that were not annotated for the 3D MRI.
It is assumed that the voxels that were not annotated by experts were correctly not annotated.

\section{Experiments}\label{sec:experiments}
In this section, we compare three partially supervised methods and one fully supervised method using the fetal brain 3D MRI dataset described in section~\ref{sec:dataset}.

\begin{figure}[tb]
    \centering
    \includegraphics[width=\columnwidth]{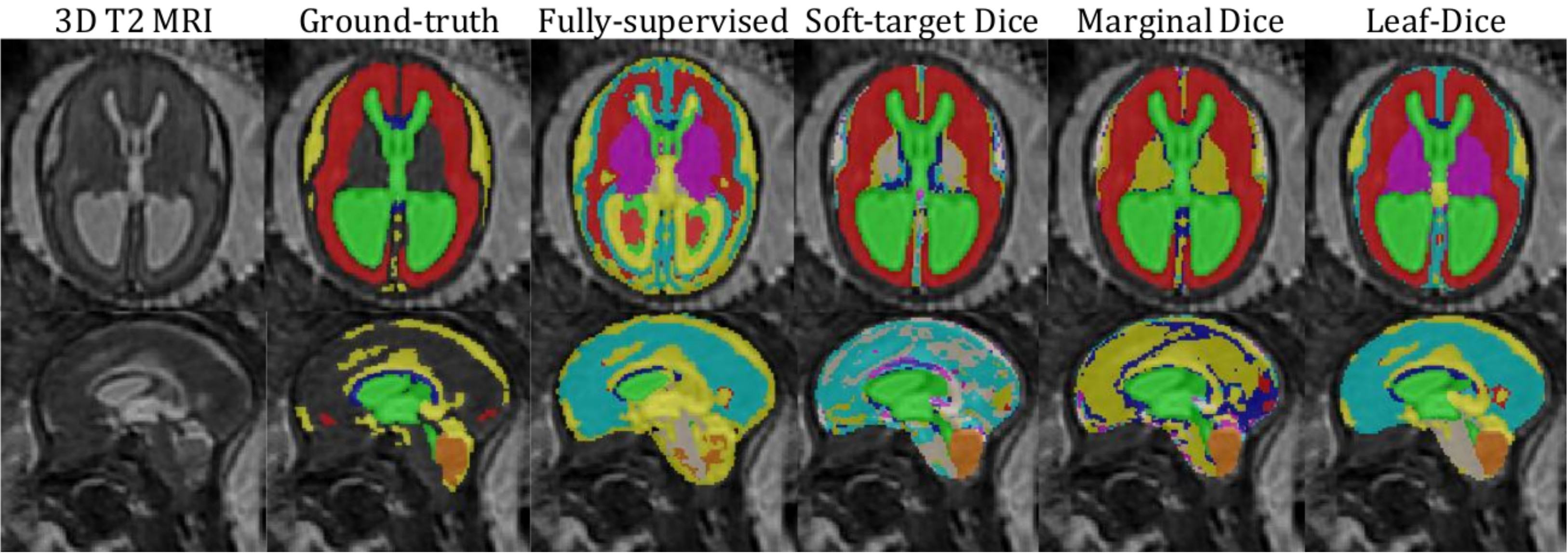}
    \caption{Qualitative comparison on an open spina bifida case.
    Only the proposed Leaf-Dice loss provides satisfactory segmentations for all tissue types.
    }
    \label{fig:qualitative_results}
\end{figure}

\subheading{Deep Learning Pipeline.}
We use a 3D U-Net~\cite{cciccek20163d} architecture with $5$ levels, 
leaky ReLU and instance normalization~\cite{ulyanov2016instance}.
For training and inference, the entire volumes were used as input of the 3D U-Net after padding to $144\times160\times144$ voxels.
%
For each method of Table~\ref{tab:models_results}, an ensemble of $10$ 3D U-Net is used.
Using an ensemble of models 
makes the comparison of segmentation results across methods less sensitive to the random initialization of each network.
Each 3D U-Net is trained using a random split of the training dataset into $90\%$ for training and $10\%$ for early stopping.
During training, we used a batch size of $3$, a learning rate of $0.001$, and Adam~\cite{kingma2014adam} optimizer with default hyperpameter values. 
The learning rate was tuned for the fully supervised learning baseline.
Random right/left flip, 
random scaling, 
gamma intensity augmentation,
contrast augmentation,
and additive Gaussian noise
were used as data augmentation during training.
Our code for the label-set loss functions and the deep learning pipeline are publicly available\footnote{\url{https://github.com/LucasFidon/label-set-loss-functions}}$^{,}$\footnote{\url{https://github.com/LucasFidon/fetal-brain-segmentation-partial-supervision-miccai21}}.

\subheading{Hardware.}
For training we used NVIDIA Tesla V100 GPUs.
Training each model took from one to two days.
For inference, we used a NVIDIA GTX 1070 GPU.
The inference for a 3D MRI takes between one and two minutes.

\subheading{Specificities of Each Method.}
Baseline 1 is trained using fully supervised learning,
the mean class Dice loss~\cite{fidon2017generalised,milletari2016v}, referred as $\mathcal{L}_{Dice}$, and the training dataset for fully supervised learning of Section~\ref{sec:dataset} (18 training volumes only).
The three other methods are trained using partially supervised learning and the training dataset for partially supervised learning of Section~\ref{sec:dataset}.
Baseline 2 is trained using the soft-target Dice loss function defined as 
$\mathcal{L}_{Dice}(\textbf{p}, \Psi_0(\textbf{g}))$, where $\Psi_0$ is defined in \eqref{eq:psi0}. 
Note that 
Baseline 2
does not satisfy the
label-set axiom~\eqref{eq:axiom}.
%
Baseline 3 is trained using the marginal Dice loss~\cite{shi2021marginal}. 
Our method is trained using the proposed Leaf-Dice loss defined in \eqref{eq:ls_dice}. 
The loss functions of Baseline~3 and our method satisfy the axiom of label-set loss functions \eqref{eq:axiom}.

\subheading{Statistical Analysis.}
We used the two-sided Wilcoxon signed-rank test.
Differences in mean values are considered significant when $p < 0.01$.

\begin{table}[t!]
	\caption{\textbf{Evaluation on the Multi-centric Testing Data (100 3D MRI).}}
	We report mean (standard deviation) for the Dice score (DSC) in percentages and the Hausdorff distance at $95\%$ (HD95) in millimeters for the eight tissue types.
    \underline{Methods underlined} are trained using partially supervised learning.
    \textbf{Loss functions in bold} satisfy the axiom of label-set loss functions.
    Best mean values are in bold.
    Mean values significantly better with $p<0.01$ (resp. worse) than the ones achieved by the fully-supervised learning baseline are marked with a $*$ (resp. a $\dag$).
	\begin{tabularx}{\columnwidth}{c *{9}{Y}}
	    \toprule
		\textbf{Model} & \textbf{Metric} & 
		\textcolor{red}{\bf WM} & \textcolor{ForestGreen}{\bf Vent} &
		\textcolor{orange}{\bf Cer} & \textcolor{Goldenrod}{\bf ECSF} &
		\textcolor{Cyan}{\bf CGM} & \textcolor{Purple}{\bf DGM} &
		\textcolor{Gray}{\bf BS} & \textcolor{blue}{\bf CC}\\
		\midrule
		Baseline 1 & DSC & 76.4 (12.5) & 72.1 (18.8) & 67.3 (28.6) & 63.9 (31.3) & 47.3 (10.9) & \textbf{72.7} (8.7) & 56.0 (27.7) & 51.6 (10.5)\\
		Fully-Supervised & HD95 & 3.3 (1.2) & 4.8 (3.7) & 4.9 (5.2) & 7.3 (7.8) & 5.3 (0.9) & \textbf{5.8} (2.1) & 9.1 (8.2) & 5.1 (3.1)\\
		\cmidrule(lr){1-10}
		\underline{Baseline 2} & DSC & $89.5^*$ (6.5) & $87.5^*$ (9.6) & $87.2^*$ (10.3) & $37.7^{\dag}$ (34.2) & $31.4^{\dag}$ (14.8) & $18.2^{\dag}$ (20.5) & $20.2^{\dag}$ (13.1) & $12.0^{\dag}$ (10.8)\\
		Soft-target Dice & HD95 & $1.7^*$ (1.3) & $1.6^*$ (2.2) & $2.2^*$ (4.5) & 7.1 (6.8) & $6.2^{\dag}$ (2.0) & $24.2^{\dag}$ (9.0) & $33.9^{\dag}$ (5.9) & $29.3^{\dag}$ (8.0)\\
		\cmidrule(lr){1-10}
		\underline{Baseline 3}~\cite{shi2021marginal} & DSC & $89.6^*$ (6.7) & $87.7^*$ (10.4) & $87.6^*$ (9.5) & $66.6^*$ (27.7) & 43.9 (15.1) & $37.8^{\dag}$ (11.3) & $39.4^{\dag}$ (16.8) & $11.1^{\dag}$ (12.3)\\
		\textbf{Marginal Dice} & HD95 & $1.7^*$ (1.3) & $1.6^*$ (2.2) & $2.4^*$ (5.4) & $6.2^*$ (7.6) & $4.4^{*}$ (1.4) & $26.7^{\dag}$ (6.7) & $33.4^{\dag}$ (6.1) & $28.7^{\dag}$ (6.3)\\
		\cmidrule(lr){1-10}
		\underline{Ours} & DSC & $\bf91.5^*$ (6.7) & $\bf90.7^*$ (8.9) & $\bf89.6^*$ (10.1) & $\bf75.3^*$ (24.9) & $\bf56.6^*$ (14.3) & 71.4 (8.6) & $\bf61.5^*$ (21.7) & $\bf62.0^*$ (10.9)\\
		\textbf{Leaf-Dice} & HD95 & $\bf1.5^*$ (1.1) & $\bf1.4^*$ (2.0) & $\bf1.7^*$ (1.8) & $\bf5.4^*$ (8.3)& $\bf3.9^*$ (1.3) & $7.3^{\dag}$ (2.3) & $\bf7.9$ (4.0) & $\bf2.9^*$ (1.5)\\
        \bottomrule
	\end{tabularx}
	\label{tab:models_results}
\end{table}

\subheading{Comparison of Fully Supervised and Partially Supervised Learning.}
The quantitative evaluation can be found in Table~\ref{tab:models_results}.
The partially supervised learning methods of Table~\ref{tab:models_results} all perform significantly better than the fully supervised learning baseline in terms of Dice score and Hausdorff distance on the tissue types for which annotations are available for all the training data of Table~\ref{tab:data}, i.e. white matter, ventricles, and cerebellum.
Some tissue types segmentations are scarce in the partially supervised training dataset as can be seen in Table~\ref{tab:data}, i.e. cortical gray matter, deep gray matter, brainstem, and corpus callosum.
This makes it challenging for partially supervised methods to learn to segment those tissue types.
Only Leaf-Dice achieves similar or better segmentation results than the fully supervised baseline for the scarce tissue types, except in terms of Hausdorff distance for the deep gray matter.
The extra-ventricular CSF is an intermediate case with almost half the data annotated.
The Leaf-Dice and the Marginalized Dice significantly outperforms the fully supervised baseline for all metrics for the extra-ventricular CSF, and the soft-target Dice performs significantly worse than this baseline.

\subheading{Comparison of Loss Functions for Partially Supervised Learning.}
The Leaf-Dice performs significantly better than the soft-target Dice for all tissue types and all metrics.
The Marginal Dice performs significantly better than the soft-target Dice in terms of Dice score for extra-ventricular CSF, cortical gray matter, deep gray matter, and brainstem.
Since the soft-target Dice loss is the only loss that does not satisfy the proposed axiom for label-set loss functions, this suggests label-set loss functions satisfying our axiom perform better in practice.

In addition, our Leaf-Dice performs significantly better than the Marginal Dice~\cite{shi2021marginal} for all metrics and all tissue types.
The qualitative results of Fig.~\ref{fig:qualitative_results} also suggest that Leaf-Dice performs better than the other approaches.
%
This suggests that using a converted fully supervised loss function, as proposed in section~\ref{sec:conversion} and in previous work~\cite{fang2020multi,roulet2019joint,shi2021marginal}, 
may be outperformed by dedicated generalised label-set loss functions.

\section{Conclusion}
In this work, we present the first axiomatic definition of label-set loss functions for training a deep learning model with partially segmented images.
We propose a generalization of the Dice loss, Leaf-Dice, that complies with our axiom for the common case of missing labels that were not manually segmented.
We prove that loss functions that were proposed in the literature for partially supervised learning satisfy the proposed axiom.
In addition, we prove that there is one and only one way to convert a loss function for fully segmented images into a label-set loss function for partially segmented images. 

We propose the first application of partially supervised learning to fetal brain 3D MRI segmentation.
Our experiments illustrate the advantage of using partially segmented images in addition to fully segmented images. 
The comparison of our Leaf-Dice to three baselines suggests that label-set loss functions that satisfy our axiom perform significantly better for fetal brain 3D MRI segmentation.

\subsubsection*{Acknowledgments}
This project has received funding from the European Union's Horizon 2020 research and innovation program under the Marie Sk{\l}odowska-Curie grant agreement TRABIT No 765148.
This work was supported by core and project funding from the
Wellcome [203148/Z/16/Z; 203145Z/16/Z; WT101957], and EPSRC [NS/A000049/1; NS/A000050/1; NS/A000027/1].
TV is supported by a Medtronic / RAEng Research Chair [RCSRF1819\textbackslash7\textbackslash34].

%
%
%
\bibliographystyle{splncs04.bst}
\bibliography{main.bib}

\newpage
\begin{center}
\textbf{\Large --- Supplemental Document ---\\Label-set Loss Functions for Partial Supervision:\\
Application to Fetal Brain 3D MRI Parcellation}\\[.6cm]
Lucas Fidon,$^{1}$ Michael Aertsen,$^{2}$ Doaa Emam,$^{4,5}$ Nada Mufti,$^{1,3,4}$ \\Fr\'ed\'eric Guffens,$^{2}$ Thomas Deprest,$^{2}$ Philippe Demaerel,$^{2}$ \\Anna L. David,$^{3,4}$ Andrew Melbourne,$^{1}$ S\'ebastien Ourselin,$^{1}$ \\Jan Deprest,$^{2,3,4}$ Tom Vercauteren$^{1}$\\[.3cm]
  \small ${}^1$School of Biomedical Engineering \& Imaging Sciences, King’s College London, UK\\
  ${}^2$Department of Radiology, University Hospitals Leuven, Belgium\\
  ${}^3$Institute for Women's Health, University College London, UK\\
  ${}^4$Department of Obstetrics and Gynaecology, University Hospitals Leuven, Belgium\\
  ${}^5$Department of Gynecology and Obstetrics, University Hospitals Tanta, Egypt
\end{center}

\setcounter{figure}{0}
\setcounter{table}{0}
\setcounter{page}{1}
\setcounter{section}{0}
\makeatletter

\section{Summary of mathematical notations}

\begin{itemize}
    \item $N$: number of voxels. 
    \item $\mathbf{L}$: set of final labels (e.g. white matter, ventricular system).
    We call the elements of $\mathbf{L}$ \textit{leaf-labels}.
    \item $2^{\mathbf{L}}$: set of subsets of $\mathbf{L}$
    We call the elements of $2^{\mathbf{L}}$ \textit{label-sets}.
    \item $P(\mathbf{L})$: space of probability vectors for leaf-labels.
    \item $\mathds{1}$: indicator function. 
    let A an assertion, $\mathds{1}(A)=1$ if $A$ is true and $\mathds{1}(A)=0$ if $A$ is false.
\end{itemize}

\section{Proof of equation (3)}
Let $\mathcal{L}:\,\, \mathbf{E} \xrightarrow{} \mathds{R}$, with $\textbf{E}:=P\left(\mathbf{L}\right)^N \times \left(2^{\mathbf{L}}\right)^N$.
%
%
Let us first remind that we have defined label-set loss functions as the losses $\tilde{\mathcal{L}}$ that satisfy the axiom
\begin{equation}
    \forall \textbf{g} \in \left(2^{\mathbf{L}}\right)^N,\,\forall \textbf{p}, \textbf{q} \in P(\textbf{L})^N,\quad
    \Phi(\textbf{p}; \textbf{g}) =\Phi(\textbf{q}; \textbf{g})
    \implies 
    \tilde{\mathcal{L}}(\textbf{p},\textbf{g}) = \tilde{\mathcal{L}}(\textbf{q}, \textbf{g})
    \label{eq:supp_labels_set_losses}
\end{equation}

\subsubsection*{First implication.}
Let us assume that 
$\forall (\textbf{p},\textbf{g})\in \textbf{E},\,\, \mathcal{L}(\textbf{p}, \textbf{g}) = \mathcal{L}(\Phi(\textbf{p}; \textbf{g}), \textbf{g})$.
Let $\textbf{g} \in \left(2^{\mathbf{L}}\right)^N$ and $\textbf{p}, \textbf{q} \in P(\textbf{L})^N$,
such that $\Phi(\textbf{p}; \textbf{g}) =\Phi(\textbf{q}; \textbf{g})$.
We have
\begin{equation*}
    \mathcal{L}(\textbf{p},\textbf{g}) = \mathcal{L}(\Phi(\textbf{p}; \textbf{g}), \textbf{g})
= \mathcal{L}(\Phi(\textbf{q}; \textbf{g}), \textbf{g})
= \mathcal{L}(\textbf{q},\textbf{g}) 
\,\,\blacksquare
\end{equation*}
%
%

\subsubsection*{Second implication.}
Let us assume that $\mathcal{L}$ is a label-set loss function.
%
%
We will use the following lemma which is proved at the end of this paragraph. 
\begin{lemma}
For all $\textbf{g} \in \left(2^{\mathbf{L}}\right)^N$,
the function $\Phi(\cdot\,; \textbf{g}): \textbf{p} \mapsto \Phi(\textbf{p}; \textbf{g})$
is idempotent, 
i.e. $\forall \textbf{p}\in P(\mathbf{L})^N,\, 
\Phi(\Phi(\textbf{p}; \textbf{g}); \textbf{g}) = \Phi(\textbf{p}; \textbf{g})$.
\label{lemma:idempotent}
\end{lemma}
Let $(\textbf{p},\textbf{g}) \in \textbf{E}$.
By Lemma~\ref{lemma:idempotent},
$
\Phi(\Phi(\textbf{p}; \textbf{g}); \textbf{g}) = \Phi(\textbf{p}; \textbf{g}).
$
Therefore, by applying \eqref{eq:supp_labels_set_losses} for $\textbf{p}$ and $\textbf{q}=\Phi(\textbf{p}; \textbf{g})$ we eventually obtain
$
\mathcal{L}(\textbf{p},\textbf{g}) = \mathcal{L}(\Phi(\textbf{p}; \textbf{g}), \textbf{g})
$
$\blacksquare$

\subsubsection*{Proof of Lemma 1}
Let $\textbf{g}= (g_i) \in \left(2^{\mathbf{L}}\right)^N$ and $\textbf{p}=(p_{i,c}) \in P(\textbf{L})^N$.
Let us denote $\Phi(\textbf{p};\textbf{g})= (\tilde{p}_{i,c})$
and $\Phi(\Phi(\textbf{p};\textbf{g});\textbf{g}) = (\vardbtilde{p}_{i,c})$.
For all $i, c$,
if $c \not \in g_i$, 
$\vardbtilde{p}_{i,c} = \tilde{p}_{i,c}$,
and if $c \in g_i$,
$
\vardbtilde{p}_{i,c} = \frac{1}{|g_i|}\sum_{c' \in g_i}\tilde{p}_{i,c'} = 
\frac{1}{|g_i|}\sum_{c' \in g_i}\left(\frac{1}{|g_i|}\sum_{c'' \in g_i}p_{i,c''}\right) = \tilde{p}_{i,c}
$
$\blacksquare$

\section{Proof that the leaf-Dice is a label-set loss function}
Let $\textbf{g} \in \gset$ be a partial segmentation that takes its value in a partition of $\mathbf{L}$ of the form
$\{\mathbf{L}'\} \cup \left\{\{c\}\,|\,c \in \mathbf{L}\setminus\mathbf{L}'\right\}$ with $\mathbf{L}' \subsetneq \mathbf{L}$
that contains all the labels of the regions of interest that were not manually segmented.
Let $\textbf{p} \in \pset$.

\begin{equation}
\begin{split}
    \mathcal{L}_{Leaf-Dice}(\textbf{p}, \textbf{g}) 
    &=
    1 - 
    \frac{1}{|\mathbf{L}|} 
    \sum_{c \in \mathbf{L}} 
    \frac{2 \sum_i \mathds{1}(g_i =\{c\})\,p_{i,c}}{
    \sum_i \mathds{1}(g_i =\{c\})^{\alpha}
    + \sum_{i} p_{i,c}^{\alpha}
    +\epsilon
    }\\
    &=
    1 - 
    \frac{1}{|\mathbf{L}|} 
    \sum_{c \in \mathbf{L}\setminus\mathbf{L}'} 
    \frac{2 \sum_i \mathds{1}(g_i =\{c\})\,p_{i,c}}{
    \sum_i \mathds{1}(g_i =\{c\})^{\alpha}
    + \sum_{i} p_{i,c}^{\alpha}
    +\epsilon
    }
\end{split}
\label{eq:supp_LSDice}
\end{equation}
because the terms of the sum for the leaf-labels $c \in \mathbf{L}'$ have a numerator equal to $0$.

To prove that $\mathcal{L}_{Leaf-Dice}$ is a label-set loss function it is sufficient to prove that 
$
\mathcal{L}_{Leaf-Dice}(\textbf{p},\textbf{g})=\mathcal{L}_{Leaf-Dice}(\Phi(\textbf{p;\textbf{g}}),\textbf{g}).
$
For all voxel i and leaf-label c
\begin{equation*}
    \left[\Phi(\textbf{p;\textbf{g}})\right]_{i,c} =
    \left\{
    \begin{array}{cc}
        \frac{1}{|g_i|}\sum_{c' \in g_i} p_{i,c'} & \textup{if}\,\, c \in g_i \\
        p_{i,c} & \textup{if}\,\, c \not \in g_i
    \end{array}
    \right.
    =
    \left\{
    \begin{array}{cc}
        \frac{1}{|\mathbf{L}'|}\sum_{c' \in \mathbf{L}'} p_{i,c'} & \textup{if}\,\, c \in g_i \,\,\textup{and}\,\, g_i = \mathbf{L}' \\
        p_{i,c} & \textup{if}\,\, c \in g_i \,\,\textup{and}\,\, g_i =\{c\}\\
        p_{i,c} & \textup{if}\,\, c \not \in g_i
    \end{array}
    \right.
\end{equation*}
where the second equality comes from the partition.

We observe that $\left[\Phi(\textbf{p;\textbf{g}})\right]_{i,c} \neq p_{i,c}$ only when $c \in \mathbf{L}'$.
Using \eqref{eq:supp_LSDice} we obtain
$
\mathcal{L}_{Leaf-Dice}(\textbf{p},\textbf{g})=\mathcal{L}_{Leaf-Dice}(\Phi(\textbf{p;\textbf{g}}),\textbf{g})\,
\blacksquare
$

\section{Proof of equation (7)}
Using previous results and the fact that $\mathcal{L}_{partial}$ is here defined as a converted fully-supervised loss function,
$\mathcal{L}_{partial}$ is a label-set loss function if and only if 
for all 
$
(\textbf{p},\textbf{g}) \in \pset \times \gset
$
\begin{equation*}
    \mathcal{L}_{partial}(\textbf{p}, \textbf{g}) 
    = \mathcal{L}_{partial}(\Phi(\textbf{p}; \textbf{g}), \textbf{g})
    = \mathcal{L}_{fully}(\Phi(\textbf{p}; \textbf{g}), \Psi(\textbf{g}))
\end{equation*}
As a result, we only need to prove that if $\mathcal{L}_{partial}$ is a label-set loss function then $\Psi=\Psi_0$.
Let us suppose that $\mathcal{L}_{partial}$ is a label-set loss function.
Therefore, for all 
$
\textbf{g} \in \gset,\,\, 
\mathcal{L}_{fully}(\Phi(\Psi(\textbf{g}); \textbf{g}), \Psi(\textbf{g}))
= \mathcal{L}_{partial}(\Phi(\Psi(\textbf{g}); \textbf{g}), \textbf{g})
= \mathcal{L}_{partial}(\Psi(\textbf{g}), \textbf{g})
= \mathcal{L}_{fully}(\Psi(\textbf{g}), \Psi(\textbf{g})).
$
Using the unicity of the minima of $\mathcal{L}_{fully}$ we can therefore conclude that 
$
\forall \textbf{g} \in \gset,\,\, \Psi(\textbf{g}) = \Phi(\Psi(\textbf{g}); \textbf{g}).
$
For all $\textbf{g},i,c$,
\begin{equation*}
    \left[\Phi(\Psi(\textbf{g});\textbf{g})\right]_{i,c} =
    \left\{
    \begin{array}{cc}
        \frac{1}{|g_i|}\sum_{c' \in g_i} \left[\Psi(\textbf{g})\right]_{i,c'} & \textup{if}\,\, c \in g_i \\
        \left[\Psi(\textbf{g})\right]_{i,c} & \textup{if}\,\, c \not \in g_i
    \end{array}
    \right.
    =
    \left\{
    \begin{array}{cc}
        \frac{1}{|g_i|} & \textup{if}\,\, c \in g_i \\
        0 & \textup{if}\,\, c \not \in g_i
    \end{array}
    \right.
    = \left[\Psi_0(\textbf{g})\right]_{i,c}
\end{equation*}
where in the second inequality we have used
$
\forall i,c,\,\, [\Psi(\textbf{g})]_{i,c} > 0 \implies c \in g_i\,\, \blacksquare
$

\section{Relation to the marginal Dice loss}
Let $P=\{\mathbf{L}_j\}$ a partition of $\mathbf{L}$ and $\textbf{g}\in \gset$ with its values in $P$.
For all $c \in \mathbf{L}$, we denote $j(c)$ the unique index such that $c \in \mathbf{L}_{j(c)}$.

Using the definitions of $\Psi_0$, $\Phi$, and the partition $P$, we obtain
\begin{equation*}
    \begin{split}
        \mathcal{L}_{Dice}\left(\Phi(\textbf{p}; \textbf{g}), \Psi_0(\textbf{g})\right)
        &=
        1 - \frac{1}{|\mathbf{L}|} \sum_{c \in \mathbf{L}} 
        \frac{
            \sum_i
            \Psi_0(\textbf{g})_{i,c}
            \Phi(\textbf{p}; \textbf{g})_{i,c}
            }{
            \sum_i \Psi_0(\textbf{g})_{i,c}
            +
            \sum_i \Phi(\textbf{p}; \textbf{g})_{i,c}
            }\\
        &=
        1 - \frac{1}{|\mathbf{L}|} \sum_{c \in \mathbf{L}} 
        \frac{
            \sum_i
            \left(
            \frac{1}{|\mathbf{L}_{j(c)}|}\mathds{1}\left(g_i = \mathbf{L}_{j(c)}\right)
            \right)
            \left(
            \frac{1}{|\mathbf{L}_{j(c)}|}\sum_{c'\in \mathbf{L}_{j(c)}}p_{i,c'}
            \right)
            }{
            \sum_i
            \frac{1}{|\mathbf{L}_{j(c)}|}\mathds{1}\left(g_i = \mathbf{L}_{j(c)}\right)
            +
            \sum_i
            \frac{1}{|\mathbf{L}_{j(c)}|}\sum_{c'\in \mathbf{L}_{j(c)}}p_{i,c'}
            }\\
        &=
        1 - \frac{1}{|\mathbf{L}|} \sum_{c \in \mathbf{L}} 
        \frac{
            \left(\frac{1}{|\mathbf{L}_{j(c)}|}\right)^2
            \sum_i
            \mathds{1}\left(g_i = \mathbf{L}_{j(c)}\right)
            \left(\sum_{c'\in \mathbf{L}_{j(c)}}p_{i,c'}\right)
            }{
            \frac{1}{|\mathbf{L}_{j(c)}|}
            \left(
            \sum_i \mathds{1}\left(g_i = \mathbf{L}_{j(c)}\right)
            +
            \sum_i \sum_{c'\in \mathbf{L}_{j(c)}}p_{i,c'}
            \right)
            }\\
        &=
        1 - \frac{1}{|\mathbf{L}|} \sum_{c \in \mathbf{L}} 
        \frac{1}{|\mathbf{L}_{j(c)}|}
        \frac{
            \sum_i
            \mathds{1}\left(g_i = \mathbf{L}_{j(c)}\right)
            \left(\sum_{c'\in \mathbf{L}_{j(c)}}p_{i,c'}\right)
            }{
            \sum_i \mathds{1}\left(g_i = \mathbf{L}_{j(c)}\right)
            +
            \sum_i \sum_{c'\in \mathbf{L}_{j(c)}}p_{i,c'}
            }\\
    \end{split}
\end{equation*}
By grouping the terms of the first sum with respect to the label-sets in $P$
\begin{equation*}
    \begin{split}
        \mathcal{L}_{Dice}\left(\Phi(\textbf{p}; \textbf{g}), \Psi_0(\textbf{g})\right)
        &=
        1 - \frac{1}{|\mathbf{L}|} \sum_{\mathbf{L}' \in P} \sum_{c \in \mathbf{L}'} 
        \frac{1}{|\mathbf{L}_{j(c)}|}
        \frac{
            \sum_i
            \mathds{1}\left(g_i = \mathbf{L}_{j(c)}\right)
            \left(\sum_{c'\in \mathbf{L}_{j(c)}}p_{i,c'}\right)
            }{
            \sum_i \mathds{1}\left(g_i = \mathbf{L}_{j(c)}\right)
            +
            \sum_i \sum_{c'\in \mathbf{L}_{j(c)}}p_{i,c'}
            }\\
        &=
        1 - \frac{1}{|\mathbf{L}|} \sum_{\mathbf{L}' \in P} \sum_{c \in \mathbf{L}'} 
        \frac{1}{|\mathbf{L}'|}
        \frac{
            \sum_i
            \mathds{1}\left(g_i = \mathbf{L}'\right)
            \left(\sum_{c'\in \mathbf{L}'}p_{i,c'}\right)
            }{
            \sum_i \mathds{1}\left(g_i = \mathbf{L}'\right)
            +
            \sum_i \sum_{c'\in \mathbf{L}'}p_{i,c'}
            }\\
        &=
        1 - \frac{1}{|\mathbf{L}|} \sum_{\mathbf{L}' \in P}
        \frac{
            \sum_i
            \mathds{1}\left(g_i = \mathbf{L}'\right)
            \left(\sum_{c'\in \mathbf{L}'}p_{i,c'}\right)
            }{
            \sum_i \mathds{1}\left(g_i = \mathbf{L}'\right)
            +
            \sum_i \sum_{c'\in \mathbf{L}'}p_{i,c'}
            }\\
    \end{split}
\end{equation*}
because for all $\mathbf{L}' \in P$ and all $c \in \mathbf{L}'$, $\mathbf{L}_{j(c)}=\mathbf{L}'$.

The right hand-side of the last equality is the marginal Dice loss~\cite{shi2021marginal} up to the multiplicative constant $\frac{1}{|\mathbf{L}|}$ $\blacksquare$

\end{document}